\documentclass[12pt,preprint]{aastex}
\usepackage{psfig}

\slugcomment{Accepted to the P.A.S.P.}
\shortauthors{Howell et al.}
\shorttitle{WZ Sge Superoutburst Photometry}

\begin{document}

\title{Multi-Color Photometry of the 2001 Superoutburst \\ of WZ Sagittae}

\author{Steve B. Howell}
\affil{WIYN Observatory and National Optical Astronomy Observatory,\\
950 N. Cherry Ave., Tucson, AZ 85719; howell@noao.edu}
\author{Arne A. Henden}
\affil{
Universities Space Research Association, U.S. Naval Observatory \\
Flagstaff Station, P.O. Box 1149, Flagstaff, AZ 86002; aah@nofs.navy.mil
}
\author{Arlo U. Landolt}
\affil{Department of Physics and Astronomy, Louisiana State University, \\
Baton Rouge, LA 70803; landolt@baton.phys.lsu.edu}
\and
\author{Courtney Dain}
\affil{IGPP, University of California, Riverside, CA 92521}

\begin{abstract}
We present the results of U-band and multi-color photometry during the
2001 superoutburst of WZ Sagittae. Our 10 nights of 
U-band photometry span the time interval from
HJD 2452118 to 2452197 while our multi-color observations 
range from HJD 2452115 to 2452197.
The U-band light curves are generally in agreement with other
datasets obtained during the superoutburst showing highly modulated light
early on, rebrightenings, and superhumps of similar shape and period (except
during the rebrightening peak).
One of our multi-color datasets fortuitously covers the first rebrightening
and allows determination of the accretion disk color temperature
before, during, and after the event.
It is seen that the rebrightening is a change from a neutral disk 
(T$\sim$7000K) to an ionized disk (T$\sim$10000K) and back again. 
We develop a simple limit cycle 
model for this behavior which approximately predicts the semi-periodic 
timescale observed for the rebrightenings. We discuss our results
in relation to accretion disk structure during superoutburst.
\end{abstract}

\section{Introduction}
WZ Sagittae is often considered the quintessential short period
dwarf nova, especially for the class of objects called TOADS 
(Tremendous Outburst Amplitude Dwarf novae) which only have
infrequent (yearly to decadal) superoutbursts
(see Howell et al., 1995).
WZ Sagittae is the brightest TOAD at minimum light (V=15.0),
has the longest intra-outburst timescale (20-30 years),
is the closest cataclysmic variable at 43.5 pc (Harrison et al., 2004),
and has one of the shortest known orbital periods (P=81.6 min).

Study of dwarf novae during superoutburst provide astronomers with
information ranging from the binary orbital period to a mass estimate
for the secondary star via detailed study of the period, shape, and
time scale of so-called superhumps visible in the light curves
(e.g., Patterson et al., 2002 and references therein).
Superhumps are periodic hump-like modulations observed in the
photometric signature of the star during superoutburst with
periods a few percent different from that of the binary orbital period.
Additionally, the morphology of the superoutburst itself, often
combined with multi-wavelength observations, can provide a direct
measurement of the properties of the accretion disk and its behavior
during the outburst (e.g., Howell et al., 1999).

Superoutbursts are often observed via world-wide observer
networks with the majority of observations being made by small
telescopes and amateur astronomers. Their
telescopes are generally equipped with CCDs and observations are
routinely obtained in ``white light". These unfiltered CCD images,
when convolved with quantum efficiencies typical of the CCDs used,
produce ``pink light" observations. Only one TOAD, WX Cet,
has multi-color photometric superoutburst observations (Howell et al.,
2002) from which the authors concluded that the superhump period was
grey in the optical and the observed period agreed with that
determined solely by white light observations.

During August 2001, WZ Sagittae erupted in superoutburst 11 years
prior to prediction and was observed by essentially every professional
telescope
available (both on the ground and in orbit) in addition to hundreds
of amateurs around the world. A summary of the massive ground-based 
observational
campaign obtaining CCD photometric observations and a detailed
interpretation of their meaning is presented in Patterson et al. (2002).

During this same time period, we collected multi-color photometric
measurements throughout the superoutburst as well as the first
set of U-band time series observations for WZ Sagittae during
a superoutburst. We present these data and interpret them in relation
to other photometric observations of WZ Sagittae during superoutburst
as well as relating the observations to accretion disk structure and
the formation and cause of the semi-periodic rebrightenings.

\section{Observations}
Photometric observations were obtained as time series data sets in U-band
and as single measurements in U, B, V, R, and I.
Tables 1 and 2 present observing logs of the U and multi-color observations
respectively.

\subsection{$U$-Band Observations}
Ten nights of U-band photometry were obtained at the US Naval Observatory's
Flagstaff Station (NOFS) between the beginning of WZ Sagittae's superoutburst
and its return to a normal minimum state.
Observations were obtained on the 1.0-m telescope at NOFS using the
SITe/Tektronix 1024x1024 CCD with an 11x11 arcmin field of view.
The comparison stars listed in Henden and Landolt (2001)
were used to set the nightly zero points (with star 2/C being the primary
comparison) but no transformation to
the standard system was performed. The NOFS U-band filter is
similar to one suggested by Bessell (1995)
which gives reasonable transformations for normal-colored stars.
Since the same comparison star(s) were used for all time-series
nights and the color difference between WZ Sge and the comparison
star(s) is not extreme, the small transformation error involved
in not transforming the data only results in a minor zeropoint
adjustment ($\sim$0.02 mag), lost in the chaotic noise 
of the intrinsic variation
of WZ Sge and not affecting the light curve analysis that follows.
For each night, standard CCD bias subtraction and flat fielding were
performed, followed by aperture photometry and differential
photometry using the inhomogeneous ensemble photometry technique
described in Honeycutt (1992).

Observations began on HJD 2452118 (the sixth day of the
superoutburst) and ended on HJD 2452197, outburst day 85.
The exposure times ranged from 20 to 60 seconds with time series
durations from 0.8 hours to 7.4 hours.
Table 1 lists the mean U magnitude and the duration of each time series.
	
\subsection{$UBVRI$ Observations}

Multi-color photoelectric data were obtained for WZ Sagittae's superoutburst
in U, B, V, R and I filters on the CTIO 1.5 meter telescope.
Ten nights of data were collected beginning
on HJD 2452115, day 4 of the superoutburst, and ending on HJD 2452197
or outburst day 85 (Table 2). These observations were tied to Landolt (1992)
and provide two overlapping days
(day 35 and 85) of data with the U-band observations from the Naval
Observatory.

All the CTIO observations used a 14.0 arcsecond diameter diaphragm
and from HJD 2452115 through 2452145, 10 second
exposure times were used
for all filters (resulting in tens of thousands of counts per
integration) while the remaining observations used 40 second exposure times
providing 10,000 counts in U and more in the remaining filters. 
The sky was photometric
throughout the CTIO run (with the exception of marginally photometric
conditions on HJD 2452194). All the observations were guided, thus WZ Sge
remained centered in the diaphragm at all times and no contamination occurred
from the close by (10" east) red companion star. The final
transformed magnitude and color indices are better than one percent
in all cases except U-B which is about $\pm$0.015. 

In addition, seven sets of transformed 
multi-color photometric observations were obtained
at the US Naval Observatory using the instrument and telescope setup described
above. These are also listed in Table 2.

\section{Analysis}
\subsection{U-Band Observations}
Figures 1 and 2 present our U-band time series observations (except for the
two short runs)
and show a variety of structures and modulations. 
Observations from early in the outburst (day 6) show a well defined
orbital modulation evident in similar light curves shown in
Patterson et al. (2002, See their Fig. 2). Superhumps became evident
on HJD 2452137 (around day 24/25)
as seen in Figure 1 (lower panels). Superoutburst day 28
(HJD 2452140) in Figure 2 shows the most consistent superhump
light curve peaking continuously at an approximate U magnitude of 12.
Three nights later, on HJD 2452143 (outburst day 31) the light curve
changed drastically as it dropped slightly (a few tenths) and with
no evidence of superhumps. This particular light curve occurred
at the peak of the first rebrightening (see discussion below).
Forty days later (HJD 24521097) modulations
returned to the light curve, essentially those typical of minimum light.

We searched for periodic behavior in each night's U-band observation
using the phase dispersion minimization (PDM) routine of Stellingwerf (1978).
Our results are presented in Table 3. Our errors for each determined period 
are fairly large (3 to 5 minutes) as each dataset is either rather short
or the light curve contains non-periodic modulations with only outburst days 25
and 28 being good representations of familiar superhumps while outburst day
85 revealed a nearly minimum orbital period..
These two days also provide similar looking results and equal periods
(within the statistics) compared with the pink light photometry
presented in Patterson et al. (2002).

\subsection{Multi-Color Photometry}
Using our multi-color observations during the 
superoutburst, we can explore the physical state in the accretion disk. Figures 3 and 4 show the three epochs during the
superoutburst for which we have multi-color data. We note here that 
during the superoutburst (V brighter than $/sim$14), 
the optically thick (boundary layer) accretion disk 
dominates the light in the (U) B through I bands (see 
Howell et al., 1999) while at minimum light, the white
dwarf contributes nearly 50\% of the optical flux (see Ciardi et al., 1998). 
The mass donor star has yet
top be directly observed, even during quiescence in the IR (Howell et 
al., 2004). The left most panel
in each plot covers days 3 to 5 of the superoutburst, the middle panel
cover days 30 to 35 (the time interval of the first rebrightening), and
the right most panel covers superoutburst days 78 to 88, essentially
when the visual outburst was over (V$\sim$14.5, about 0.5 mag above minimum).

During an outburst, the emitted flux from an accretion disk is well represented
by optically thick (blackbody) emission. This approximation scheme was first
detailed and applied to accretion disks by la Dous (1989) in which each broad
band color (i.e., U, B, etc.) provides a good approximation of the emissive
flux for successively hotter to cooler (inner to outer) single temperature
disk (see Fig. 2.8
in Warner 1995). More detailed accretion disk modeling, using stellar 
atmospheres and hydrodynamic codes has been applied today, but for our order of
magnitude discussion below, the blackbody approximation is sufficient.
Note that during (super)outburst, an optically thick approximation for the
accretion disk
emission is valid (see Fig. 3.32 in Warner 1995) but is generally 
invalid during
quiescence, at least for the low mass transfer (short orbital period) CVs (see
Fig. 2.33 in Warner 1995).

Using the above assumption that the accretion
disk emission is incandescent (i.e., optically thick) 
during the early superoutburst (days 3 to
at least 70), the photometric colors can supply us with
color temperatures for the disk emitting regions as the outburst progresses.
In 
the far right panel (near the end of the superoutburst), we can not be certain
that the observed colors, at this time,
can be attributed to optically thick material as WZ Sge's accretion disk at
minimum light has optically thin components (see Mason et al., 2001)
Table 4 lists our determined color temperatures within the two early time
intervals under the assumption that the emitted light is from optically thick
gas, that is, it can be modeled as blackbody emission.

Examination of the
early multi-color observations shows that the U-B color became redder
from superoutburst day 3 to 5, while the other colors remained fairly
constant albeit being blue. The reason that
U-B became redder is likely related to the diminished high
temperature of the
inner disk and boundary layer which rapidly drops early on in the
superoutburst and this region becomes optically thin (see Howell et al., 1999).

The middle panels in Figs 3 \& 4 show a very interesting 
time in the superoutburst as this time period spans, from start to finish, the
first rebrightening. 
The peak of the first rebrightening occurred on superoutburst day 32.6
(HJD 24521043.6; see Patterson et al., 2002) the time when all of the
colors (except U-B, see below) change to a
much bluer (hotter color temperature) value. 
The entire first rebrightening is seen to start and end near V=12.7 with redder
colors and a temperature near 6800-7500K.
During the peak of the rebrightening at V=11.2, the colors are bluer 
and we see color temperatures of 9000-10000K (in B, V, and R) but only 7700K
in R-I. The U-B color is bluer at rebrightening maximum 
but the temperatures derived from this color index are very 
high indeed and likely indicate that the emitting region may be
fully ionized throughout the rebrightening. This being the case, we can no 
longer believe in the complete validity of a determined U-B color 
temperature derived from the assumption of optically thick gas. Howell 
et al. (1999) show that during early superoutburst,
the inner disk regions are quite hot and optically thin (i.e., 
a central disk hole exists). 
The hot white dwarf surface and the boundary layer probably provide 
most of the U-band light during this time period.

Essentially all TOADs show a cooling wave effect during their superoutburst.
This event is observed as a large drop in brightness near
day 25-45 in their superoutburst light curve (e.g. Richter 1992). 
The source of this cooling wave dip in superoutburst 
light curves occurs near the inner accretion disk (Kuulkers et al., 1996). 
WZ Sagittae and EG 
Cnc before it, produced a number of semi-periodic rebrightenings during 
this ``dip" time period in place of a single large dip. We note that 
the previous two superoutbursts of WZ Sagittae, unlike this one, 
show a single large cooling wave
dip (Richter 1992; Patterson 1980) of total duration 
$\ga$9 days (1946) and 19 days (1978) while 
the total length of the 12 rebrightenings in the 2001 superoutburst 
was 23 days.  While of a longer duration, the ``by eye" integral under
all the rebrightenings in the 2001 superoutburst is nearly the same
as the area in the single cooling dip during 1978.
We note in Fig. 2, that at the start of the cooling wave dip
(outburst day 27) no superhumps are present in the U-band.
It is tempting to try to draw a conclusion that the
33 year interval up to the 1978 superoutburst produced a single large 
cooling wave dip while the shorter 21 year interval up to the 
2001 superoutburst led to numerous rebrightenings during this same 
time period.

The cause of the drop in light output during the dip 
is a quenching of the disk outburst by a 
cooling (density) wave of mainly neutral material moving through the 
disk. If enough disk material is piled up ahead of the wave, it can 
cause a back pressure and thus reflect the cooling wave 
causing a rebrightening, thus ending the dip. 
It is assumed that these cooling wave dips are 
not seen (as dramatically at least) in normal dwarf novae outbursts
as the amount of material involved in
the outburst (moving through the disk) is too small. In order to see 
if a thermal limit cycle may be at work in the accretion 
disk and cause the semi-periodic rebrightenings, we develop here a toy 
model. Based on our multi-color observations we see that the color 
temperature crosses the hydrogen 
ionization boundary during the rebrightening. The simple starting assumption 
for the rebrightening is that a volume of neutral hydrogen gas is heated 
to $\sim$10000K and becomes hotter and brighter (bluer) as well as soon
becoming fully ionized. It then
cools (mostly by line radiation with some expansion) and returns to 
neutral gas at which time it will be cooler and fainter. 

During the first rebrightening, 
the color temperature derived for most of the disk (Table 4),
is seen to oscillate from below the hydrogen ionization
temperature to above it at the peak of the event. 
The R-I color is an exception and probably indicates that while 
the outer accretion disk temperature changes as well (higher during the peak)
the outer disk never becomes fully ionized.
The behavior of the colors and the derived color
temperatures during this interval
suggests that the accretion disk regions which participate in
the general decline of the superoutburst
(assuming pure hydrogen composition)
start and end mainly neutral, but become ionized near 
the peak of the rebrightening. 
This type of behavior and the multiple rebrightenings 
lead us to suggest a model of the accretion disk 
during the rebrightenings which operates as a limit cycle.

\section{Simple Limit Cycle Model}

Since the rebrightenings are oscillatory 
with a ``period" near 1.5 to 2 days (see Fig.1 in Patterson et al., 2002) we 
desired to determine if some characteristic timescale within the disk 
during superoutburst might cause such an effect. 

We can establish the luminosity of the start (end) and peak of the first
rebrightening during the superoutburst as we now know the
distance to WZ Sagittae (43.5 pc, Harrison et al, 2004).
Using the V magnitude and standard filter bandpass, we find the values to be
(start \& end) $8.0 \times 10^{28}$ and 
(peak) $2.5 \times 10^{30}$ ergs/sec respectively, in good agreement with the 
predicted superoutburst luminosity at this time based on the 
model presented in Howell et al. (1999). Using the luminosity and 
color temperature values before (and after) and during the rebrightening, 
we can calculate the radius of the assumed spherical 
optically thick emitting region. 
Taking average color temperature values from 
Table 4 to be T$_{cool}$=7000K and T$_{hot}$=10,000K respectively,
the emitting region
radii are then found to be R$_{cool}$=$2.15 \times 10^{8}$ cm 
and R$_{hot}$=$5.9\times 10^{8}$ cm. 
These values are again in agreement with the emitting sizes determined for 
TOADs during superoutburst when allowing for a time period of $\sim$30 
days past peak brightness (Howell et al., 1999). 
Note here that we have made use of the simplifying assumption that each
broad band color provides an estimate of a single temperature 
for regions within the accretion disk.
The true situation is complex with a mix of temperatures, but
since we do not resolve the accretion disk, our photometric measurements average
over the entire accretion disk area (i.e., temperature structure) 
visible at each epoch, leading to a mean temperature estimate within each 
color. 

The total cooling rate, G, in a gas can be expressed as 
$$G = L_R + L_{FF} + L_C,$$ where $R$ = radiative, $FF$ = free-free, 
and $C$= collisional cooling (Osterbrock 1974, \S3.7). Following this 
argument and taking typical values for number densities in TOAD accretion 
disks (N$_e$$\sim10^{10}$ to $10^{14}$ cm$^{-3}$),
assuming that each hydrogen atom will (eventually) lose near 13.6 eV 
($2.18 \times 10^{-11}$ ergs) during the cooling process,
and the participating disk volume is the spherical volume discussed above
(we could use a cylinder here, V = $\pi~R_{hot}^2h$ 
where h is set to 0.1R$_{hot}$, but for our toy model the difference is 
of no consequence) we find that the effective cooling 
rate will be in the range of $2.2 \times 10^{21}$
to $2.2 \times 10^{25}$ ergs/sec. 
Integrating the Saha equation\footnote{The Saha equation, when solved for the
ratio of ionized to total hydrogen, tells us that for densities 
near N$_e$$\sim10^{12}$, HII/HI=0 at 6000K and below but reaches 1.0 by 12,000K.
The 5000K range in between shows a sharp, non-linear rise from all
neutral to all
ionized with the halfway point (HII/HI=0.5) being near 6400K (for our low
density limit) and near 9200K (for our high density limit).} 
as one crosses the ionized to neutral 
temperature boundary (noting that the cooling required to return to neutral
conditions drops rapidly as the temperature drops)  
and using the assumed shrinking emitting volume, 
as R$_{hot}$ changes to R$_{cool}$, we find that the cooling times required
(over the range of likely disk densities and assuming the range of $G$ above) 
range from a few hundred seconds (at the 
high density end) up to nearly 3 days for the low density values. 

During the superoutburst, with a hotter expanded disk,
we might expect the densities to be better
represented by the lower density values. The time scale determined for these
densities (a few days) matches well that of the nearly periodic
1.5-2 day rebrightenings observed in 
WZ Sge. Thus, our toy model
may provide a natural mechanism for the limit cycle causing the 
rebrightenings; semi-periodic cooling wave - reflected heating wave 
oscillations causing some portion of the disk gas to modulate
across the sharp temperature boundary between mainly neutral
to mainly ionized hydrogen. The reason for a single
cooling wave dip,
or its replacement by semi-periodic rebrightenings, may be some change
in the amount of material involved in the superoutburst and/or the
local disk conditions at the time. 
Note that in WZ Sge the accretion disk radius is approximately 
equal to the diameter of Jupiter (see Fig. 12 in Skidmore et al., 2001) 
while the emitting regions determined above 
are Earth-sized, making our use of an assumed spherical
emitting volume within the accretion disk fairly justified. This 10:1
emitting area relationship (disk to varying emitting region) 
does not seem out of line to be able to
account for the $\sim$1 magnitude (2.5 times) rebrightenings.
 
\section{Conclusion}

We find that our U-band time series photometry generally agrees in 
light curve shape and
period (both orbital modulations and superhumps) with the more detailed 
pink light results presented in Patterson et al. (2002). Our multi-color results
show that the superoutburst has a blue nature to it early on (related to the
high temperatures produced) and returns to a blue color in U-B and B-V near the
end due to the emergence of the hot white dwarf. Additionally, we find that the
rebrightenings appear to be a transition in a local disk volume from mostly
neutral conditions to fully ionized gas and back. A toy model based on a
cooling wave-temperature limit cycle provides a time scale consistent with
observations.

Our limit cycle model is a subset of the typical dwarf nova accretion disk 
outburst limit cycle.
Perhaps the hydrogen ionization limit cycle operates within accretion disks 
on various physical scale lengths leading to 
many different observed phenomena such as the semi-periodic 
modulations which we call disk rebrightenings. 
This type of cycle may also be the cause for the observed rapid outbursts 
in the ER UMa stars and similar type oscillations 
at late times in classical novae outbursts. 

\acknowledgements
The authors wish to thank the referee, Michael Richmond, for his timely 
and well though out review leading to a much improved presentation.
AUL thanks the staff at Cerro Tololo Inter-American Observatory
for their usual superb support, and acknowledges support from
NSF grant AST 00-97895.

\newpage

\begin{center}
{\bf Figure Captions}
\end{center}

\figcaption[]{
U-band time series light curves for WZ Sge during superoutburst
days 6, 7, 24, and 25. Note the large orbital modulation in days 6 and 7 and 
the fully developed U-band superhumps on day 25. Both light curve features are
in agreement with the observations presented in Patterson et al., (2002).}

\figcaption[]{
U-band time series light curves for WZ Sge during superoutburst
days 27, 28, 31, and 85. Note the apparent weakening of the superhumps 
in the U-band on day 27 caused by a steep decline in the flux over the
time series. This day started the cooling wave dip phase. 
Superoutburst day 31 (lower left panel) shows the U-band light
curve at the peak of the first rebrightening during which time no superhumps
are apparent. This is in contrast to the light curves presented in Figure 5 of
Patterson et al. (2002), but see our Figure 4.
}

\figcaption[]{Our V, U-B and B-V photometry for three selected time regions
during the superoutburst. The light curve shapes and V magnitude values are in
agreement with previously published values. Note the first rebrightening
shown in the middle panel. 
During the first rebrightening, the U-band light 
(mainly from the inner disk region which is likely
to be optically thin at this time) does not participate in the 
red to blue color change observed in the other colors. This
condition likely explains the lack of superhumps observed during 
this time interval (see Fig. 2).
Note the blue color of WZ Sge both early in the superoutburst and 
during the late decline.}

\figcaption[]{V-R, R-I, and V-I photometry for the same three 
time intervals presented
in Figure 3. Note the blue color early on but a red (accretion disk dominated) 
color during the late decline. During the first rebrightening, we again note a
significant red to blue color change.
However, the R-I and V-I (due to the I value) color temperature does not
reach that for full ionization during the rebrightening peak (See Table 4).}

\newpage

\begin{deluxetable}{cccc}
\tablenum{1}
\tablewidth{5.8in}
\tablecaption{U Band Time-Series Photometry}
\tablehead{
 \colhead{HJD(2452100+)}
 & \colhead{Outburst Day}
 & \colhead{Mean U Mag (Range)}
 & \colhead{Duration (hr)}
}
\startdata
\hline
18 &		6 &		8.25 (+/-.25) &  2.88  \\
19 &		7 &		8.45 (+/-.11) &  3.12 \\
36 &		24 &		10.09 (+/-.12) & 4.08 \\
37 &		25 &		10.28 (+/-.16) & 5.04 \\
38 &		26 &		-- 	&	0.840 \\
39 &		27 &		12.1 (+.2/-.3) & 7.44 \\
40 &		28 &		12.26 (+.16/-.3) & 7.20 \\
43 &		31 &		10.65 (+.35/-.2) & 6.72 \\
52 &		40 &		11.15 (+/.15)	 & 1.39 \\
97 &		85 &		13.28 ( +.12/-.18) & 4.90 \\
\hline
\enddata
\end{deluxetable}{}

\newpage
\begin{deluxetable}{lccccccl}
\tablenum{2}
\tablewidth{6.0in}
\tablecaption{Multi-Color Photometry}
\tablehead{
 \colhead{HJD(2452100+)}
 & \colhead{V}
 & \colhead{B-V}
 & \colhead{U-B}
 & \colhead{V-R}
 & \colhead{R-I}
 & \colhead{V-I}
 & \colhead{Observer\tablenotemark{a}}
}
\startdata
\hline
15.6745 & 8.269 & -0.137 & -0.962 &   --   &   --   &   --   & AL \\
15.6774 & 8.218 & -0.134 & -0.944 & -0.009 & -0.011 & -0.023 & AL \\   
15.7310 & 8.315 & -0.120 & -0.922 & -0.014 & -0.027 & -0.044 & AL\\
16.6514 & 8.553 & -0.104 & -0.819 &   --   &   --   &   --   & AL \\
16.6543 & 8.638 & -0.088 & -0.811 & +0.001 & +0.032 & +0.027 & AL \\
16.6578 & 8.741 & -0.090 & -0.878 & +0.037 & +0.031 & +0.063 & AL \\
16.6614 & 8.768 & -0.116 & -0.838 & +0.002 & +0.008 & +0.006 & AL \\
16.6650 & 8.772 & -0.112 & -0.833 & -0.003 & +0.008 &  +0.001 & AL \\
16.6684 & 8.746 & -0.105 & -0.827 & -0.021 & -0.010 & -0.035 & AL \\
16.6720 & 8.661 & -0.100 & -0.807 & -0.003 & +0.016 & +0.009 & AL \\
16.6753 & 8.650 & -0.092 & -0.797 & -0.002 & +0.015 & +0.009 & AL \\
16.7421 & 8.706 & -0.090 & -0.802 & -0.014 & +0.003 & -0.015 & AL \\
16.7455 & 8.621 & -0.106 & -0.898 &  0.000 & +0.009 & +0.005 & AL \\
16.7491 & 8.571 & -0.107 & -0.916 & +0.001 & +0.003 &  0.000 & AL \\
16.7525 & 8.544 & -0.111 & -0.898 & -0.014 & -0.001 & -0.020 & AL \\
16.7559 & 8.487 & -0.104 & -0.900 & -0.005 & +0.004 & -0.004 & AL \\
16.7593 & 8.487 & -0.107 & -0.872 & +0.011 & +0.007 & +0.013 & AL \\
16.7628 & 8.555 & -0.086 & -0.818 & +0.011 & +0.020 & +0.026 & AL \\
16.7661 & 8.646 & -0.077 & -0.782 & +0.015 & +0.021 & +0.030 & AL \\
18.7611  & 8.977 & -0.037 &  -0.691 & -0.042 & +0.055 & +0.013 & AH \\
37.9152  & 10.839 & +0.121 &  -0.576 & +0.034 & +0.134  & +0.168 & AH \\
39.9312  & 12.914 & +0.074 &  -0.934 & +0.225 & +0.261  & +0.486 & AH \\
42.6691 & 12.826 & +0.267 & -0.930 & +0.249 & +0.328 & +0.579 & AL \\
43.6181 & 11.213 &  +0.001 & -0.871 & +0.092 & +0.101 & +0.198 & AL \\
43.6610 & 11.227 & -0.002 & -0.868 & +0 089 & +0.085 & +0.180 & AL \\
44.5794 & 12.120 & +0.048 & -0.871 & +0.135 & +0.189 & +0.323 & AL \\
44.5954 & 12.126 & +0.042 & -0.949 & +0.130 & +0.171 & +0.300 & AL \\
44.6533  & 12.136 &  +0.087 & -0.797 &  +0.106 & +0.126 & +0.232 & AH \\
44.6635 & 12.116 & +0.069 & -0.859 & +0.142 & +0.164 & +0.305 & AL \\
44.7306  & 12.216 &  +0.145 & -0.790 &  +0.105 & +0.221 & +0.326 & AH \\
45.5823 & 12.591 & +0.283 & -0.796 & +0.243 & +0.290 & +0.528 & AL \\
45.7501  & 11.792 &  +0.266 & -0.544 &  +0.148 & +0.209 & +0.683 & AH \\
53.8371  & 11.298 &  +0.087 & -0.589 & -- & -- & -- & AH \\
91.5322 & 14.171 & -0.064 & -1.096 & +0.188 & +0.279 & +0.474 & AL \\
94.5353 & 14.221 & -0.048 & -1.125 & +0.128 & +0.189 & +0.338 & AL \\
95.5335 & 14.251 & -0.084 & -1.130 & +0.173 & +0.135 & +0.329 & AL \\
97.5263 & 14.236 & -0.024 & -1.140 & +0.092 & +0.194 & +0.267 & AL \\
\hline
\enddata
\tablenotetext{a}{AL = Arlo Landolt; AH = Arne Henden}
\end{deluxetable}{}

\newpage

\begin{deluxetable}{ccc}
\tablenum{3}
\tablewidth{4.5in}
\tablecaption{Measured U Band Periods}
\tablehead{
 \colhead{HJD(2452100+)}
 & \colhead{Outburst Day}
 & \colhead{Period ($\sigma$) (min)}
}
\startdata
\hline
18	&	6	&	78 +3.5/-13.1	\\	
19	&	7	&	82 +7.2/-9.6   \\
36	&	24	&	none		\\	
37	&	25	&	81.6 +5.3/-5.1 \\
38	&	26	&	none		\\	
39	&	27	&	85.2 +2.25/-7.78 \\
40	&	28	&	82 +5.08/-6.3  \\
43	&	31	&	none           \\
52	&	40	&	none           \\
97	&	85	&	86 +2.92/-3.5 \\
\hline
\enddata
\end{deluxetable}{}
 
\newpage
 
\begin{deluxetable}{cccccc}
\tablenum{4}
\tablewidth{6.2in}
\tablecaption{Color Temperatures (K)}
\tablehead{
 \colhead{HJD interval (2452100+)}
 & \colhead{U-B}
 & \colhead{B-V}
 & \colhead{V-R}
 & \colhead{R-I}
 & \colhead{V-I}
}
\startdata
\hline
15.5-16	& 25000 & -- & -- & -- & -- \\
16.5-17 & 19000 & -- & -- & -- & -- \\
15.5-17 & -- & 11400 & 10500 & 9500 & 9600 \\
42-43   & -- & 7500 & 7500 & 6250 & 6800 \\
43-45   & 21000-25000 & 9000 & 9000-10500 & 7300-7700 & 7800-9000 \\
45-46   & 19000 & 7500 & 7500 & 6450 & 6800 \\ 
\hline
\enddata
\end{deluxetable}{}

\newpage
\begin{figure}
\plotone{fig1.ps}
\end{figure}

\newpage
\begin{figure}
\plotone{fig2.ps}
\end{figure}

\newpage
\begin{figure}
\plotone{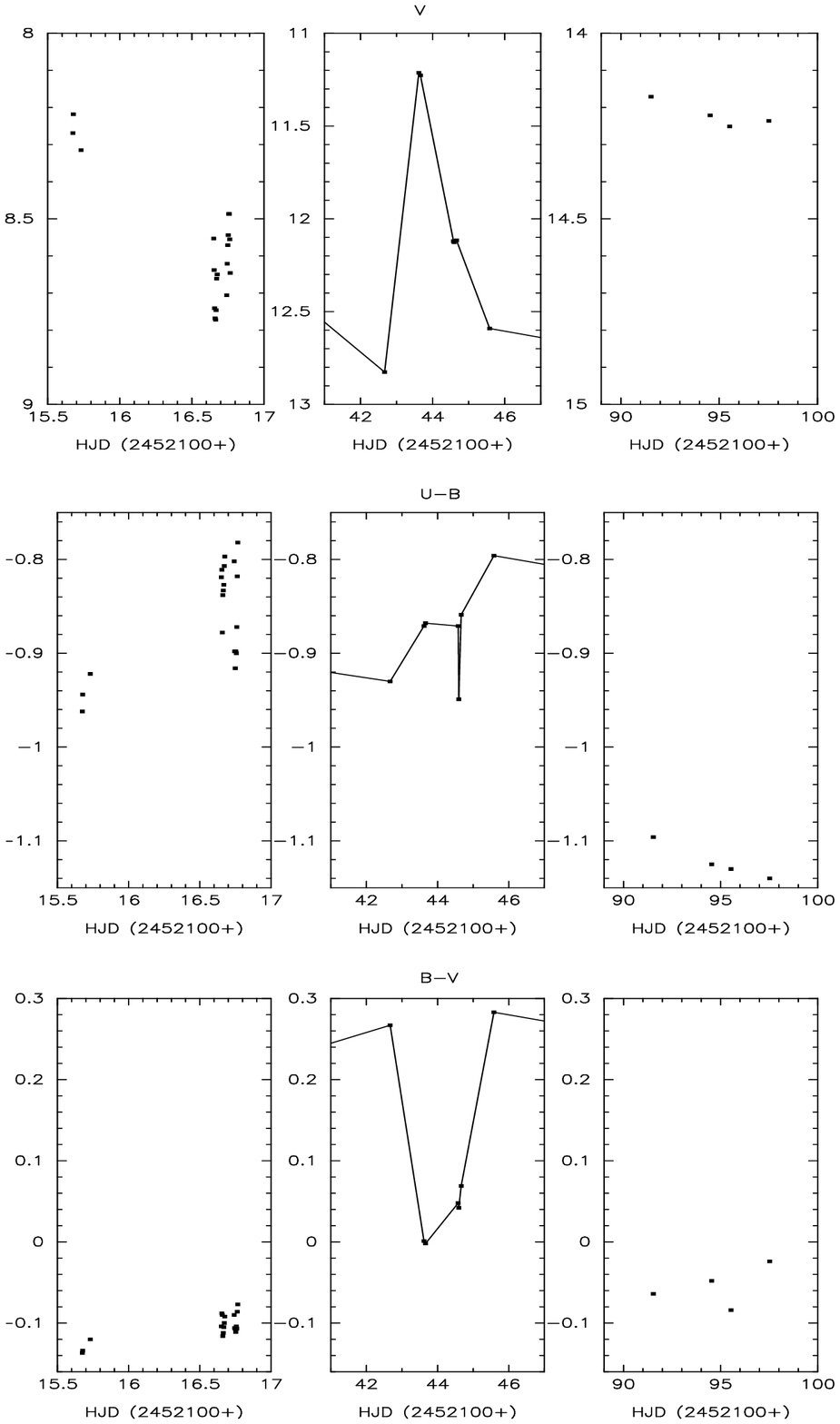}
\end{figure}

\newpage
\begin{figure}
\plotone{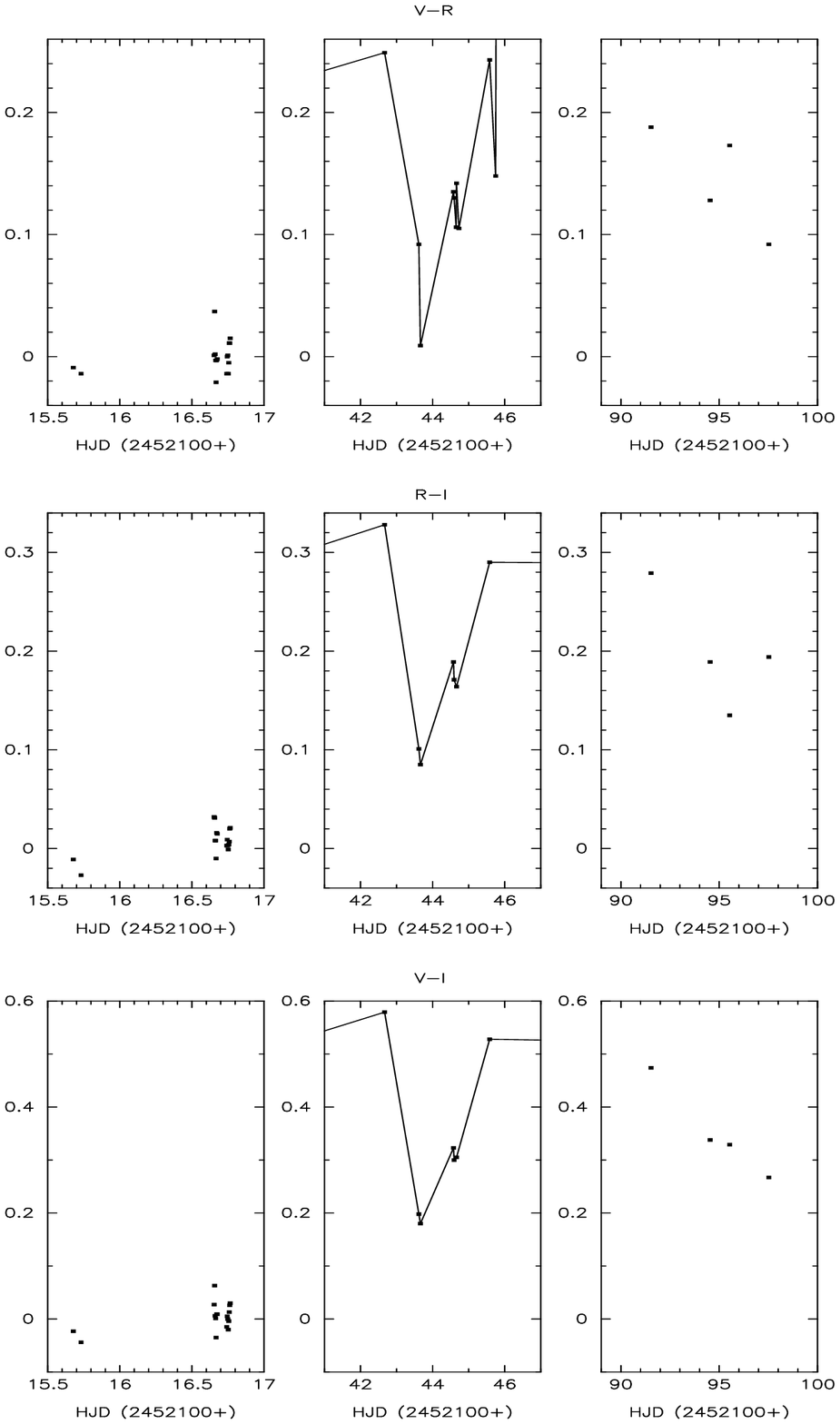}
\end{figure}

%
%
%
%
%

\end{document}